\documentclass[conference]{IEEEtran}
\usepackage{booktabs} 
\usepackage{cite}
\usepackage{subfig}
\usepackage{amsmath,amssymb,amsfonts}
\usepackage{algorithmic}
\usepackage{graphicx}
\usepackage{textcomp}
\usepackage{xcolor}
\graphicspath{ {./images/} }
\def\BibTeX{{\rm B\kern-.05em{\sc i\kern-.025em b}\kern-.08em
    T\kern-.1667em\lower.7ex\hbox{E}\kern-.125emX}}

\usepackage{float}
\usepackage{flushend} 
\begin{document}

\title{A Dynamic Approach to Load Balancing in Cloud Infrastructure: Enhancing Energy Efficiency and Resource Utilization\\}

\author{\IEEEauthorblockN{Shadman Sakib}
\IEEEauthorblockA{\textit{Department of Computer Science} \\
\textit{Missouri State University}\\
Springfield, USA\\
ss4587s@missouristate.edu}
\and
\IEEEauthorblockN{Ajay Katangur}
\IEEEauthorblockA{\textit{Department of Computer Science} \\
\textit{Missouri State University}\\
Springfield, USA\\
ajaykatangur@missouristate.edu}
\and
\IEEEauthorblockN{Rahul Dubey}
\IEEEauthorblockA{\textit{Department of Computer Science} \\
\textit{Missouri State University}\\
Springfield, USA\\
rahuldubey@missouristate.edu}
}

\maketitle

\begin{abstract}
Cloud computing has grown rapidly in recent years, mainly due to the sharp increase in data transferred over the internet. This growth makes load balancing a key part of cloud systems, as it helps distribute user requests across servers to maintain performance, prevent overload, and ensure a smooth user experience. Despite its importance, managing server resources and keeping workloads balanced over time remains a major challenge in cloud environments. This paper introduces a novel Score-Based Dynamic Load Balancer (SBDLB) that allocates workloads to virtual machines based on real-time performance metrics. The objective is to enhance resource utilization and overall system efficiency. The method was thoroughly tested using the CloudSim 7G platform, comparing its performance against the throttled load balancing strategy. Evaluations were conducted across a variety of workloads and scenarios, demonstrating the SBDLB's ability to adapt dynamically to workload fluctuations while optimizing resource usage. The proposed method outperformed the throttled strategy, improving average response times by 34\% and 37\% in different scenarios. It also reduced data center processing times by an average of 13\%. Over a 24-hour simulation, the method decreased operational costs by 15\%, promoting a more energy-efficient and sustainable cloud infrastructure through reduced energy consumption.

\end{abstract}

\section{Introduction}

Cloud computing has gained widespread adoption due to its scalable and flexible nature. More organizations are moving away from traditional on-premises infrastructure in favor of remote cloud solutions that offer greater cost efficiency, enhanced security, and improved accessibility \cite{li2023overview}. Despite its growth and long-term benefits, cloud adoption presents challenges, particularly in process management, legal compliance, data security, and system reliability \cite{tabrizchi2020survey}, \cite{pallathadka2022investigation}. These barriers are especially pronounced for small and medium-sized enterprises, which often lack the resources to maintain their own cloud infrastructure.

Infrastructure as a Service (IaaS) has emerged as one of the most adopted cloud models, providing scalable and cost-efficient alternatives \cite{zoting2025iaas}. In IaaS, cloud service providers (CSPs) such as AWS, Azure, IBM Cloud, and Google Cloud operate under service level agreements (SLAs), which establish performance expectations \cite{paul2018dynamic}. CSPs must also maintain Quality of Service (QoS), ensuring reliable resource allocation even as user demand grows. As cloud usage increases, load balancing becomes a critical challenge \cite{kumar2017dynamic}, essential for distributing workloads to prevent server overload, minimize latency, and ensure consistent performance.

Addressing these challenges, this research proposes a Score-Based Dynamic Load Balancer (SBDLB) to efficiently distribute workloads across virtual machines (VMs). Unlike traditional methods, SBDLB dynamically evaluates resource availability and assigns tasks using a computed score, promoting optimal resource utilization and balanced workload distribution. It also integrates a VM task threshold to prevent overloading, with each VM capped at a maximum number of concurrent tasks based on its capacity—determined through empirical testing under varying loads. SBDLB is compared against the widely used throttled load balancer, which has demonstrated strong performance in prior studies \cite{mohamed2021proposed}, \cite{le2022ita}.

Four simulation scenarios were developed to evaluate SBDLB in terms of average response time, data center processing time, and operational cost. Extensive experiments and statistical analysis, including p-value calculations, confirm the significance of performance improvements over the throttled method. Results show that SBDLB consistently outperforms the throttled load balancing strategy across all scenarios, reducing average response times by 34\% and 37\% in two key tests, and lowering data center processing times by an average of 13\%. In a separate experiment, SBDLB demonstrated its ability to achieve better performance with fewer active data centers. Furthermore, a 15\% reduction in operational costs was observed over a 24-hour simulation period, highlighting the efficiency of the method. By minimizing execution delays and avoiding resource overuse, SBDLB not only improves performance but also reduces energy consumption. This dual benefit enhances cost efficiency and contributes to more sustainable, environmentally friendly cloud computing operations.

\section{Related Work}
Data generation has surged in recent years \cite{duarte2024data}, with social media being one of the leading contributors \cite{sandhu2021big}. As more and more users access the cloud, load balancing has become crucial in managing large surges of data requests. Consequently, efficient load balancing techniques are essential to ensure optimal resource utilization and maintain high performance in the face of growing demand.
Researchers have increasingly recognized the significance of this problem, leading to extensive studies aimed at mitigating these challenges. Various approaches have been explored, including data center selection policies as well as advancements in load balancing techniques to enhance efficiency and resource utilization. 

Load balancing is said to be of two types \cite{jain2016survey}: (1) static and (2) dynamic. In static methods, system characteristics are predefined and do not consider real-time data, making it simple but inflexible. Dynamic methods, though more complex, adapt to current system status, enabling more efficient load balancing \cite{al2023fast}. Several researchers have explored variations of well-known load balancing techniques, including Round Robin \cite{gao2022improved}, Threshold \cite{rathore2016dynamic}, and Throttled \cite{le2022ita}. In a recent study, the authors implemented a priority-weighted Round Robin strategy to effectively manage incoming tasks with varying priorities \cite{katangur2022priority}. However, the Round Robin approach may lead to VM overload in scenarios where a high volume of tasks arrives concurrently, as it fails to account for the resource capacity and current load of individual VMs. Another study introduced a dual-threshold approach, where one threshold value identifies underloaded VMs and another detects overloaded VMs \cite{chowdhury2022threshold}. Both \cite{mohamed2021proposed} and \cite{le2022ita} introduce variations to the throttled approach, leading to modest improvements in response times. Additionally, both studies demonstrate significant enhancements over other load balancing methods, such as Round Robin, Active Monitoring, and Equal Load Distribution.

Given its established effectiveness and widespread use, the throttled approach serves as a natural baseline for comparison in this study, providing a benchmark against which the performance of the proposed dynamic algorithm can be evaluated. In another study, the authors randomly assigned values for task length and completion time before allocating these tasks to random VMs. The completion time was then calculated based on the characteristics of both the task and the VM. If the calculated time resulted in a SLA violation for the VM, the task was migrated to another VM \cite{shafiq2021load}. Random allocation of tasks to VMs is an inefficient strategy as it disregards the resource requirements of tasks and the available capacity of VMs.

Several nature-inspired algorithms have been applied to both data center selection policies  and load balancing. The genetic algorithm-based DC service broker policy proposed in \cite{chowdhury2023genetic} presents an innovative approach to minimizing network delays. However, it suffers from a prolonged convergence time, and since the genetic algorithm (GA) is executed only once at the start of every hour, it is not well-suited for dynamic environments requiring real-time adaptability. Another approach employs GA for load balancing by allocating tasks in bulk from a queue. However, this method presents several drawbacks, including a lack of real-time adaptability, increased waiting times for tasks, and inefficient handling of heterogeneous workloads \cite{zomaya2001observations}. 

A survey confirms that Meta-Heuristic approaches, such as ACO, Cuckoo Search, Honey Bee Optimization, and others, effectively balance cloud workloads. However, these algorithms have drawbacks in convergence rate, affecting exploration or exploitation \cite{pai2022analytical}, \cite{gokul2022cloud}. Recent studies have increasingly focused on green cloud computing, recognizing data centers as a major source of carbon emissions \cite{sriram2022green}. As companies and industries prioritize eco-friendly technological solutions, such optimizations align with broader sustainability goals, making the dynamic approach both economically and environmentally favorable \cite{demirbaga2025ecocloud}.

\section{System Model}
\subsection{Cloud Environment}
\begin{figure}[t]
    \centering
    \includegraphics[width=\columnwidth]{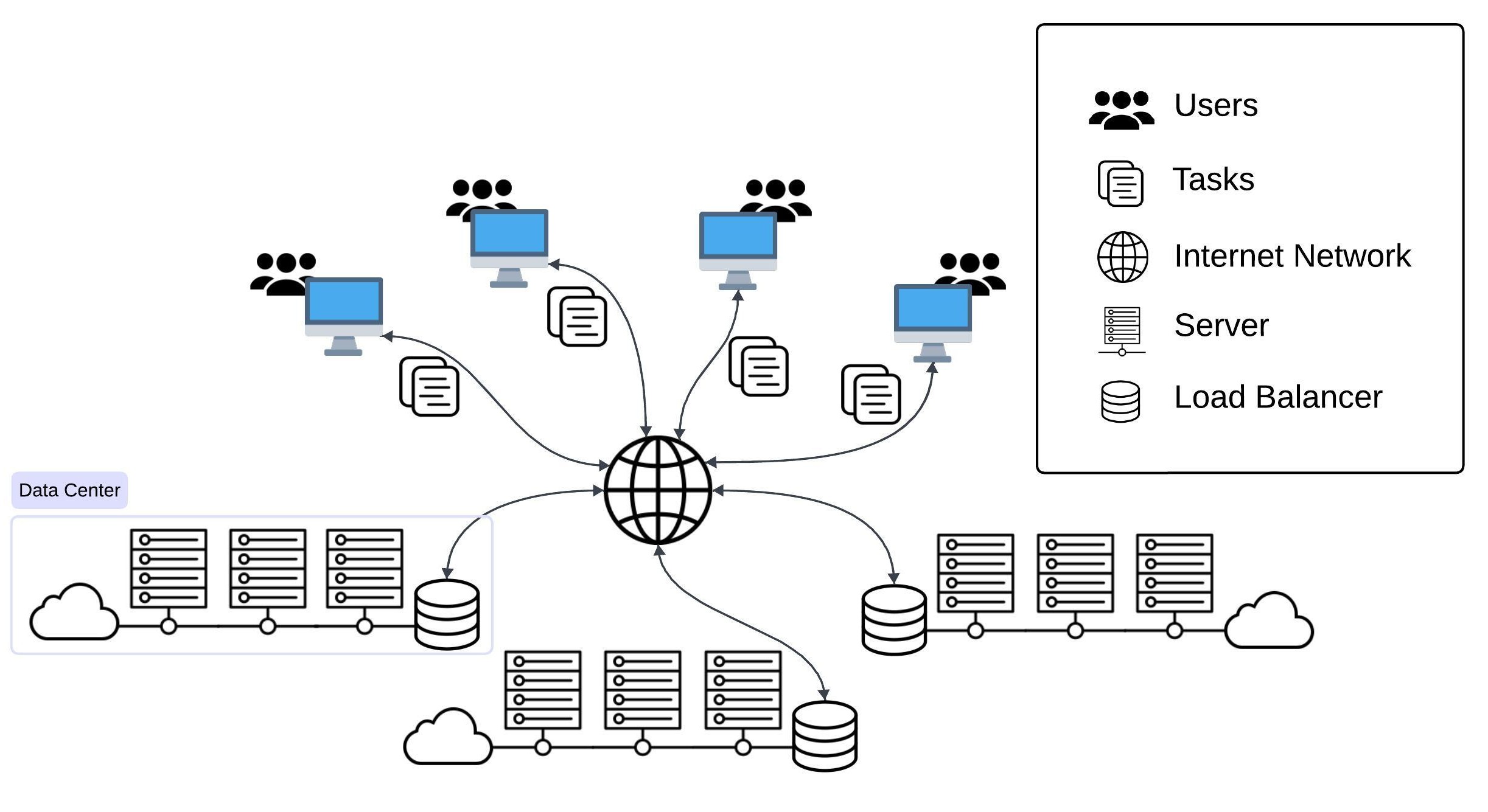}
    \caption{Simple Cloud Infrastructure}
    \label{fig:cloudpic}
\end{figure}

Figure \ref{fig:cloudpic} illustrates a simplified cloud infrastructure, where users send requests through a central gateway to distributed data centers. Load balancing ensures these requests are efficiently routed, preventing overload and maximizing resource utilization \cite{sui2019virtual}. Consider a cloud service provider with globally distributed data centers: $DC = \{ DC_{1}, DC_{2}, DC_{3}, \dots, DC_{d} \}$, each consisting of hundreds of physical machines. These machines host multiple virtual machines: $VM = \{ VM_{1}, VM_{2}, VM_{3}, \dots, VM_{n} \}$, each with different hardware configurations. When a user requests access to cloud resources, the load balancer directs the task $T$ to the most suitable data center and VM. The decision is based on factors such as resource availability, current load, and task requirements.



\subsection{System Configuration}
\label{sec:systemConfig}
For executing and evaluating the performance of the proposed load balancing technique, a simulation environment was set up using CloudSim 7G \cite{andreoli2025cloudsim}. It is a tool for modeling and testing cloud-based infrastructures and resource allocation strategies. Testing new techniques in a real cloud environment is impractical, as it may impact end-user service quality. Therefore, a reliable simulator like CloudSim is essential for tasks like scheduling and load balancing. The data centers are designed based on the specifications outlined in Table \ref{table:data_center_specs}, while the physical machines adhere to the configurations detailed in Table \ref{table:server_configs}. The VMs are created in accordance with the specifications provided in Table \ref{table:vm_specs}. For simplicity and to facilitate clearer performance comparisons, the simulation utilizes only two distinct server configurations, consistent with the methodology described by Razali \cite{razali2014virtual}. Likewise, the virtual machine setup is divided into two types each designed to meet varying computational requirements. This streamlined heterogeneous configuration enables a more focused analysis of how the load balancing algorithm manages diverse workloads. Unlike prior studies \cite{soni2014novel}, \cite{razali2014virtual}, which typically employ uniform VM configurations, this approach offers a more realistic simulation of real-world scenarios.


\begin{table}[t]
\centering
\caption{Data Center Specifications}
\renewcommand{\arraystretch}{1.2} 
\begin{tabular}{p{4.5cm} p{3cm}}
\hline
\textbf{Attribute} & \textbf{Details} \\ \hline
Architecture        & x86 \\ \hline
Operating System    & Linux \\ \hline
Virtual Machine Monitor & Xen \\ \hline
CPU Usage Cost      & \$3/sec \\ \hline
Memory Cost         & \$0.004/MB \\ \hline
Bandwidth Cost      & \$0.01/Mbps \\ \hline
Storage Cost        & \$0.0001/MB \\ \hline
\end{tabular}
\label{table:data_center_specs}
\end{table}



\begin{table}[t]
\centering
\caption{Physical Machine Specifications}
\renewcommand{\arraystretch}{1.2} 
\begin{tabular}{p{3cm} p{2cm} p{2cm}}
\hline
\textbf{Attribute} & \textbf{Type 1} & \textbf{Type 2} \\ \hline
RAM (MB)           & 1024            & 2048            \\ \hline
Storage (GB)       & 10              & 20              \\ \hline
Bandwidth (MB/s)   & 1000            & 2000            \\ \hline
Processing Cores   & 4               & 8               \\ \hline
\end{tabular}
\label{table:server_configs}
\end{table}


\begin{table}[t]
\centering
\caption{Virtual Machine Specifications}
\renewcommand{\arraystretch}{1.2} 
\begin{tabular}{p{3cm} p{2cm} p{2cm}}
\hline
\textbf{Attribute} & \textbf{Low-Spec VM} & \textbf{High-Spec VM} \\ \hline
Processing Power (MIPS) & 500 & 1000 \\ \hline
Storage (GB)          & 10  & 20   \\ \hline
RAM (MB)              & 1024 & 2048 \\ \hline
Bandwidth (MB/s)      & 1000 & 2000 \\ \hline
CPU Cores             & 1    & 2    \\ \hline
\end{tabular}
\label{table:vm_specs}
\end{table}

\section{Proposed Score-Based Dynamic Load Balancer}

\subsection{Score-Based Dynamic Load Balancer}
\label{sec:dynamic_load_balancer}

\begin{figure}[htbp]
    \centering
    \includegraphics[width=\columnwidth]{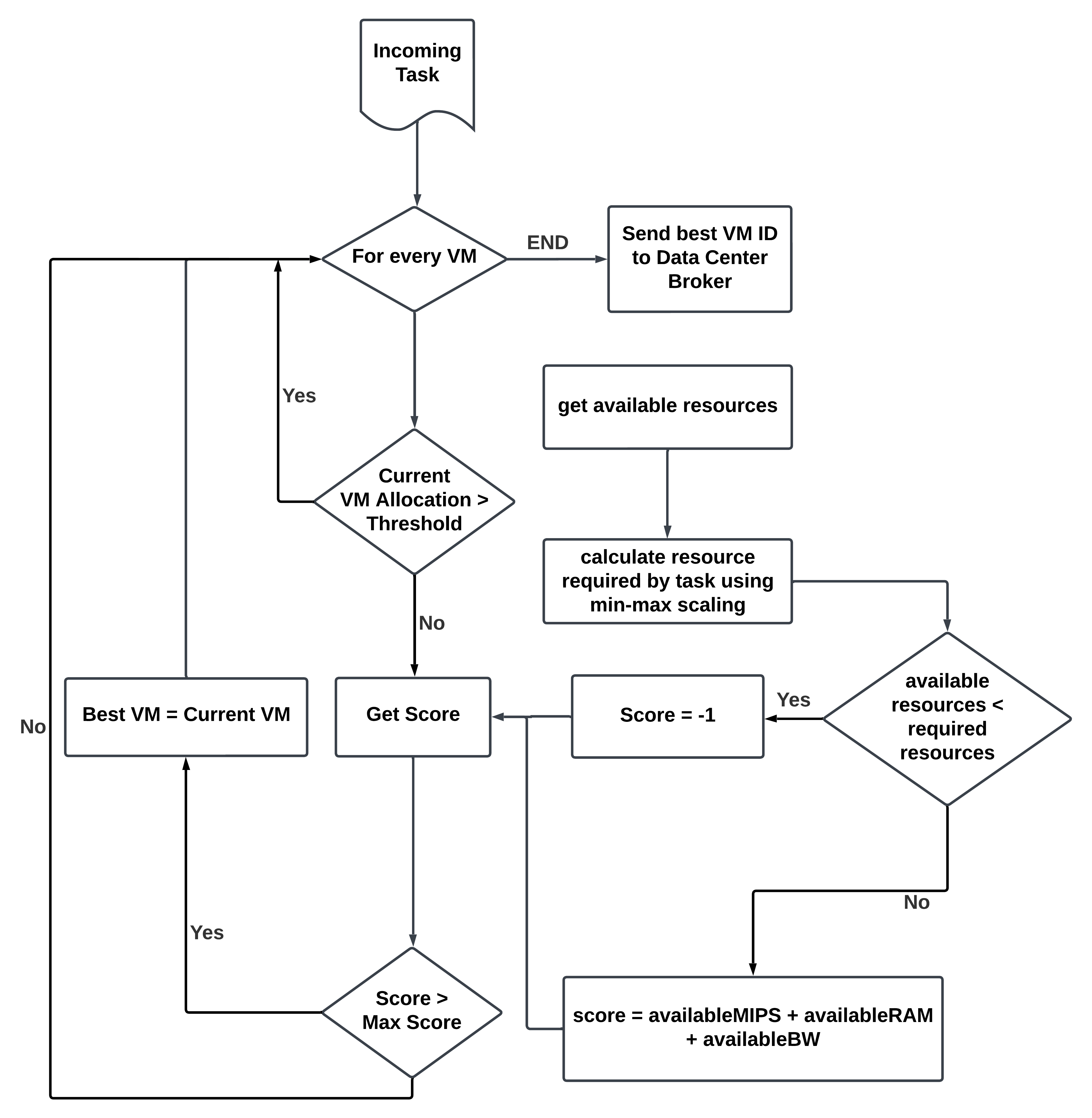}
    \caption{Score Based Dynamic Load Balancer}
    \label{fig:dynamicLB}
\end{figure}

\subsubsection{Score-Based Dynamic Load Balancer (SBDLB)}

The Score-Based Dynamic Load Balancer (SBDLB) allocates tasks by evaluating virtual machines (VMs) based on resource availability and workload to ensure efficient task distribution. Figure~\ref{fig:dynamicLB} illustrates the flow of the proposed approach. When a task arrives, the system first scans the available VMs, excluding those that exceed a predefined task threshold. For the remaining VMs, key parameters such as available CPU utilization (MIPS), RAM, and bandwidth are retrieved.

To determine the resource requirements for each incoming task, the system considers the task length, which falls within a predefined range corresponding to specific task types. The task length is then normalized using a min-max scaling technique (Equation~\ref{eq:normalization}), mapping it to the range of the available resources for each VM. This ensures that the task's resource demands are expressed in terms of the VM's available resources, such as MIPS, RAM, and bandwidth. If a VM lacks sufficient resources to accommodate the task's normalized demands, it is assigned a score of $-1$. Otherwise, a suitability score is computed by summing the available MIPS, RAM, and bandwidth as follows:
\[
\text{Score} = \text{availableMIPS} + \text{availableRAM} + \text{availableBW}
\]
The VM with the highest suitability score is selected for task allocation and passed to the Data Center Broker. Once assigned, the VM utilizes its available resources (MIPS, RAM, and bandwidth) in proportion to the task’s normalized requirements.

\begin{equation}
    y = \frac{(x - x_{\min})}{(x_{\max} - x_{\min})} \times (y_{\max} - y_{\min}) + y_{\min}
    \label{eq:normalization}
\end{equation}

Min-max scaling maps an input value \( x \) from the original range \( [x_{\min}, x_{\max}] \) to a target range \( [y_{\min}, y_{\max}] \), enabling consistent evaluation of heterogeneous task sizes and resource needs.

\subsection{Performance Metrics}
\label{sec:performance-metrics}

To evaluate the efficiency of the proposed load balancing algorithm, three core performance metrics are used: average response time, average data center processing time, and operational cost.

\subsubsection{Average Response Time}

Average response time measures the time taken to process a task from the moment it is acknowledged by the load balancer until completion. It reflects both scheduling efficiency and VM performance. Given \( n \) tasks, the average response time is calculated as:

\begin{equation*}
    \bar{R} = \frac{1}{n} \sum_{i=1}^{n} R_i
\end{equation*}

where \( R_i \) is the response time of the \( i \)-th task. Lower values indicate quicker task handling and better system responsiveness.

\subsubsection{Average Data Center Processing Time}

This metric captures the average time a data center spends processing all assigned tasks. The total processing time for a data center is given by the difference between the time the last task finishes and the time the first task starts, i.e., \( T_{\text{DC}} = T_{\text{last\_finish}} - T_{\text{first\_start}} \). To calculate the average processing time across \( n \) data centers, we sum the total processing times of each data center and divide by the number of data centers:

\begin{equation*}
    P_{\text{DC}} = \frac{1}{n} \sum_{i=1}^{n} T_{\text{DC}, i}
\end{equation*}

where \( T_{\text{DC}, i} \) is the total processing time for the \( i \)-th data center. 

\subsubsection{Operational Cost}

The operational cost of a data center is directly tied to how long it remains active for task processing. Using the total processing time calculated earlier, the cost to operate the data center is:

\begin{equation*}
    C_{\text{DC}} = T_{\text{DC}} \times CostPerSec_{\text{CPU}}
\end{equation*}

where \( \text{CostPerSec} \) is the cost of CPU usage per second. This metric is useful for comparing the cost-efficiency of different load balancing strategies.

\subsection{Setting Up Task Threshold}
\label{taskthreshold}

\begin{figure}[htbp]
    \centering
    \includegraphics[width=\columnwidth]{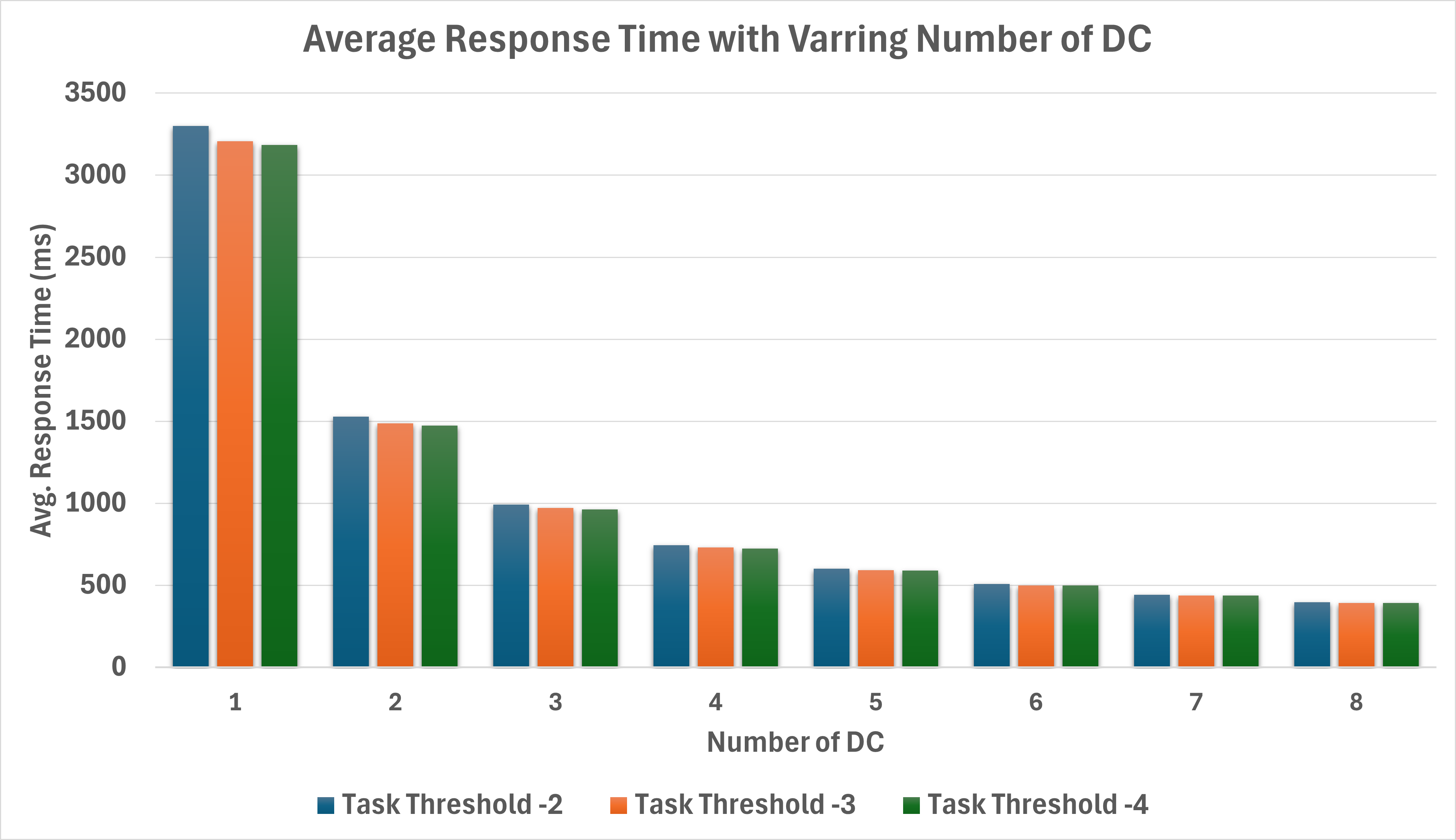}
    \caption{Average Response Time with Varying Task Threshold}
    \label{fig:alloc}
\end{figure}

To prevent VM overloading from bursts of small tasks, a task threshold was set to limit active tasks per VM. Extensive testing across workloads (Figure \ref{fig:alloc}) using 1 to 8 data centers, 2000 tasks across 250 batches, and thresholds of 2, 3, and 4, revealed that a threshold of 3 offered optimal performance. It matched the efficiency of threshold 4 while yielding lower response times than threshold 2. A threshold of 5 caused overload in single DC setups due to limited capacity.

\subsection{Task Scheduling and Load Balancing}
\begin{figure}[t]
    \centering
    \includegraphics[width=\columnwidth]{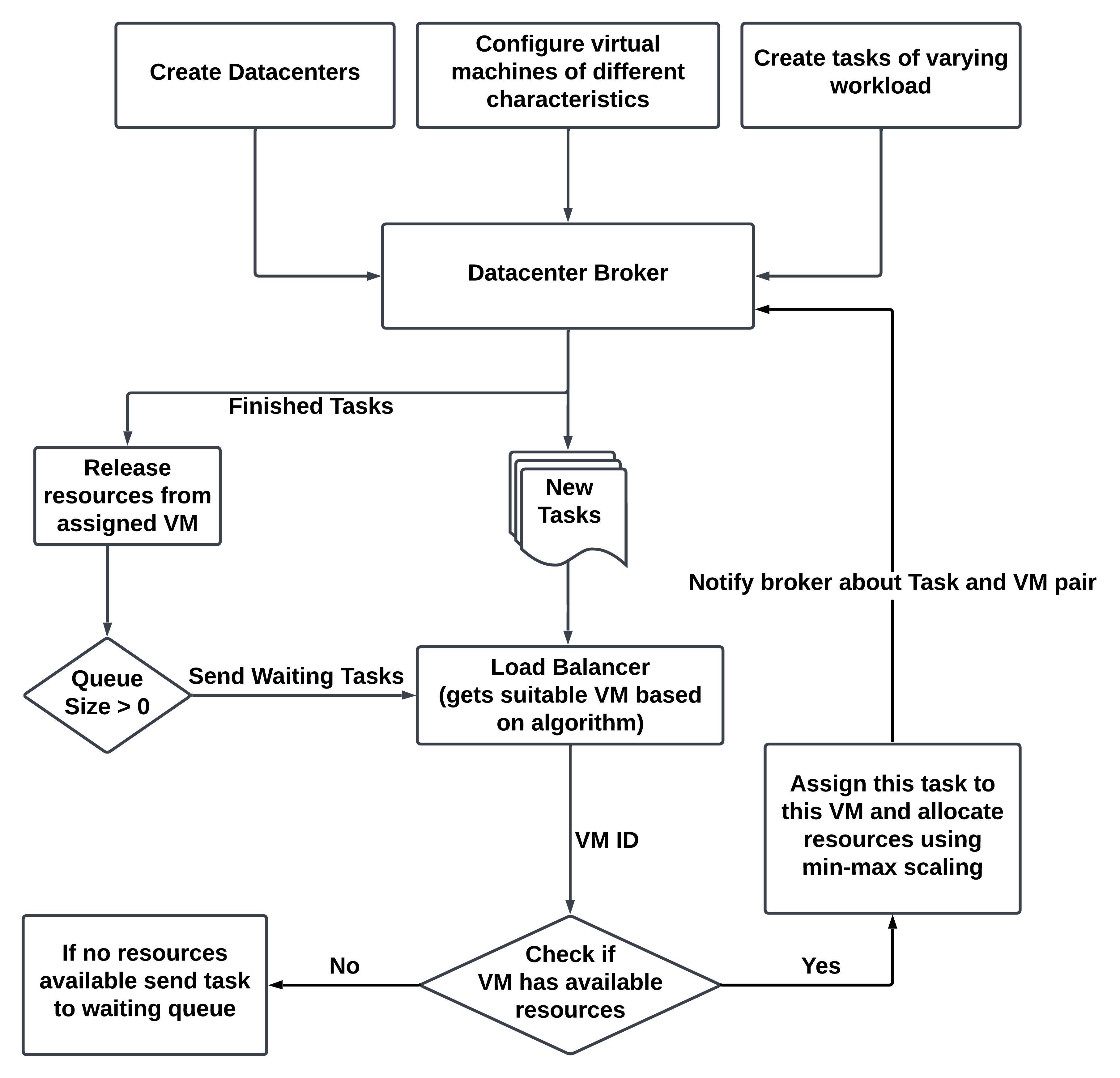}
    \caption{Task Scheduling and Load Balancing Flow in CloudSim 7G}
    \label{fig:cloudsim_flow}
\end{figure}

Figure~\ref{fig:cloudsim_flow} illustrates the workflow of the proposed simulation framework implemented in CloudSim 7G. The simulation models a cloud environment where tasks of varying complexity are dynamically scheduled to virtual machines via a data center broker and a load balancing mechanism. The simulation initializes data centers and VMs with heterogeneous configurations. Tasks are generated in batches and categorized by computational complexity. The data center broker manages task assignment using two separate load balancing strategies: (1) the Throttled Load Balancer, which distributes tasks sequentially across VMs, and (2) the proposed Score-Based Dynamic Load Balancer, which allocates tasks based on a real-time scoring mechanism. If a VM has sufficient resources, the task is assigned proportionally based on normalized task length (Equation~\ref{eq:normalization}). Longer tasks consume more resources, ensuring balanced distribution. If no VM meets the requirements, the task is queued until capacity is available. Upon task completion, resources are released and queued tasks are reassessed for execution.

\begin{figure*}[t]
    \centering
    \subfloat[Average Response Time Accross Variable Number of VMs]{%
        \includegraphics[width=\columnwidth]{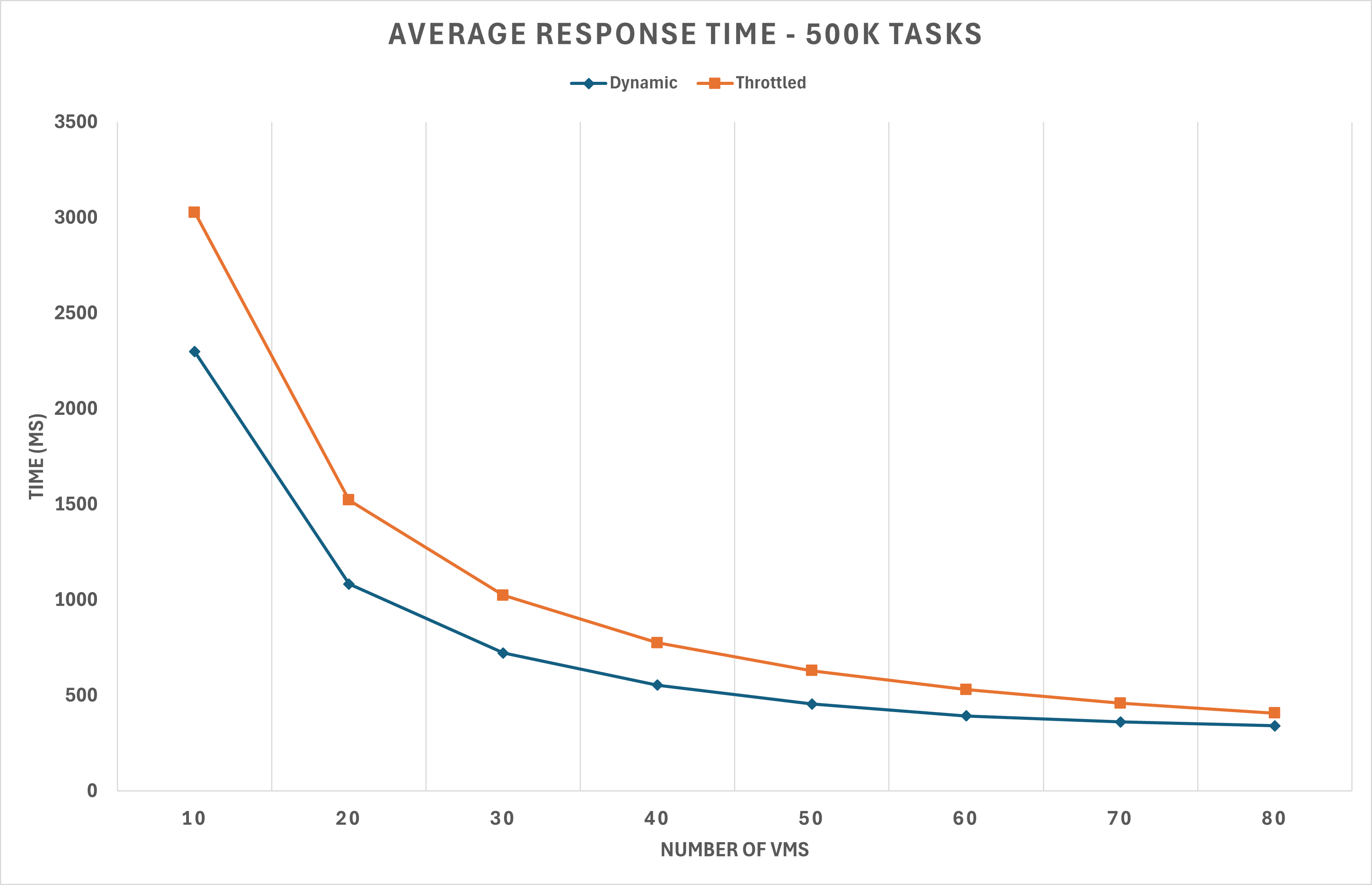}
        \label{fig:art500k}
    }
    \hfill
    \subfloat[Rate of Decrease In Average Response Time From Baseline 10 VMs]{%
        \includegraphics[width=\columnwidth]{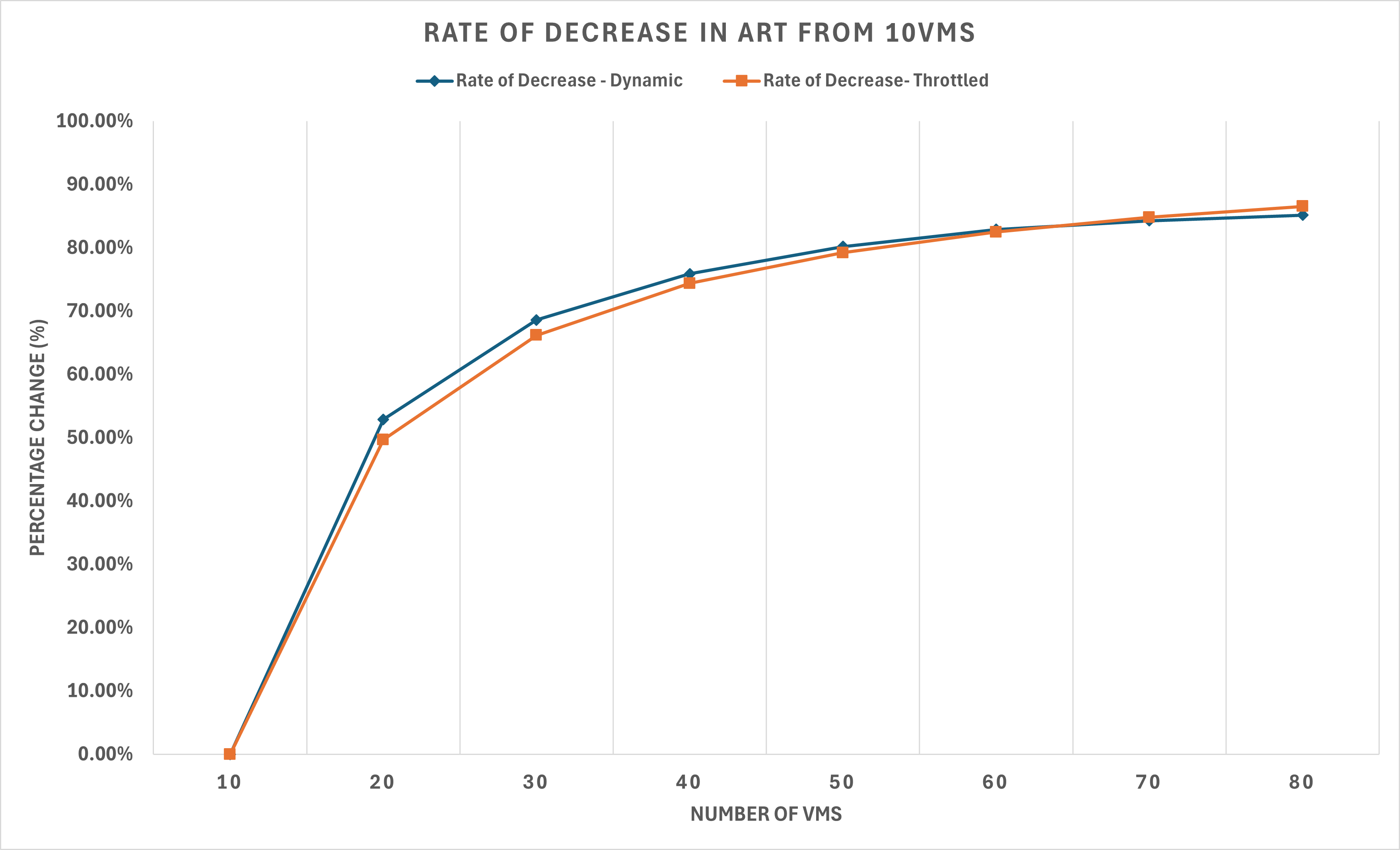}
        \label{fig:rod500k}
    }
    \caption{Performance Metric Over 10 - 80 VMs for 500K Tasks}
    \label{fig:hourly_comparison}
\end{figure*}

\section{Experimental Setup}
\subsection{Simulation Environment}
\label{subsec:task_categorization}
The simulation environment is designed to approximate the infrastructure of social media platforms. As of February 2025, Meta operates 24 data center campuses globally \cite{zhang2024}. To maintain a manageable yet realistic model, this number is scaled down by a factor of three, resulting in eight data centers used across simulation scenarios. To enhance realism, tasks are categorized into three types based on data size, complexity, and CPU requirements: \textit{Reels}, \textit{Images}, and \textit{Text Posts}. \textit{Reels} are the largest, ranging from 10 MB to 1 GB \cite{quickframe2025}, followed by \textit{Images} (1–30 MB) \cite{lamaj2024}, and \textit{Text Posts} (10–100 KB).

Task distribution follows current media consumption trends: 60\% Reels, 30\% Images, and 10\% Text. Studies show video content dominates user engagement and retention \cite{McCormick2024,Lukan2023}, while images remain crucial due to high visual processing efficiency \cite{Crackitt2018}. This breakdown reflects real-world usage and provides a practical basis for cloud task simulation in CloudSim.


\begin{table*}[t]
    \centering
    \caption{Cloud Scenarios And Analayzed Metrics}
    \begin{tabular}{|p{3.5cm}|p{4.5cm}|p{4cm}|p{4cm}|}
        \hline
        \textbf{Scenario Name} & \textbf{Description} & \textbf{Variable Factors} & \textbf{Analyzed Metrics}  \\
        \hline
        Scenario 1: VM Scalability & Evaluates how changing the number of VMs per DC affects system performance, with the total number of DCs fixed at eight &The number of virtual VMs assigned to each DC and the total volume of tasks to be processed. & Average response time \\
        \hline
        Scenario 2: Varying DC & Analyzes the effect of varying the number of data centers while keeping the number of VMs per DC constant at 60 & The number of data centers actively participating in task processing &  average response time, DC processing time, and DC operating cost \\
        \hline
        Scenario 3: Task Allocation & Examines how two load balancers distribute tasks between high-spec and low-spec virtual machines & The total number of tasks entering the system over a given period &  Task distribution across high-spec and low-spec VMs \\
        \hline
        Scenario 4: 24-Hour Variation & Assesses how effectively two load balancers distribute workloads across 60 virtual machines over a continuous 24-hour simulation period &  The timing and intensity of peak usage periods, including how many tasks are received during those high-traffic intervals & Hourly breakdowns of average response time and data center processing time and operational cost of DC for a full 24-hour period \\
        \hline
    \end{tabular}
    \label{tab:scenario_configurations}
\end{table*}

\subsection{Estimating Computational Demand by Task Type}
\label{subsec:mips-calculation}

Task size, measured in Million Instructions (MI), is determined by the workload size and the computational intensity (CI), which represents the number of CPU instructions required to process one byte of input. CI reflects the computational demand of a task and varies across task categories: lightweight tasks (e.g., text processing) typically have low CI (10–100 instructions per byte), moderate tasks (e.g., image processing) require a moderate CI (500–1,000 instructions per byte), and heavy tasks (e.g., video transcoding or compression) involve a high CI (1,000–10,000 instructions per byte). The instruction length for a given task is calculated by multiplying the data size by the task’s CI, resulting in the following formula for MI:

\[
\text{MI} = \frac{\text{Data Size (Bytes)} \times \text{CI}}{10^6}
\]

\section{Experimental Analysis}
This section presents four cloud-based simulation scenarios comparing SBDLB with throttled load balancing. Configurations and metrics are summarized in Table \ref{tab:scenario_configurations}. Experiments varied task loads from 100K to 500K in 100K steps, with a batch size of 2000. Due to space constraints, only one representative result per scenario is shown. To confirm consistency and statistical significance, p-values are included. Detailed results follow in the subsections.

\subsection{S-1: VM Scalability}

This experiment evaluated how different load balancing strategies affect system performance under varying workloads and resource configurations. It focused on assessing the scalability and effectiveness of SBDLB versus throttled-based load balancing by varying the number of VMs per data center (10 to 80) and total incoming tasks, measuring average response time as the primary metric. The system used eight active data centers. The VM range was based on preliminary results showing response time plateaus beyond 80 VMs.

SBDLB consistently outperformed throttled. At 500K tasks and across the full VM range, SBDLB achieved a 34\% lower average response time (Figure \ref{fig:art500k}). This improvement is statistically significant (p = 3.54 $\times$ 10$^{-10}$), highlighting SBDLB’s superior efficiency in handling large-scale, distributed workloads.

A key finding was that increasing VMs from 10 to 20 cut average response time by ~50\%, but gains diminished with further increases. Response time plateaued around 60 VMs (Figure \ref{fig:rod500k}), which was chosen as the standard for later experiments. This trend illustrates the point of diminishing returns: while initial VM increases yield major gains, beyond 60, added resources provide minimal benefit. This insight supports efficient infrastructure planning by balancing performance with cost.

\begin{figure}[ht]
    \centering
    \includegraphics[width=\columnwidth]{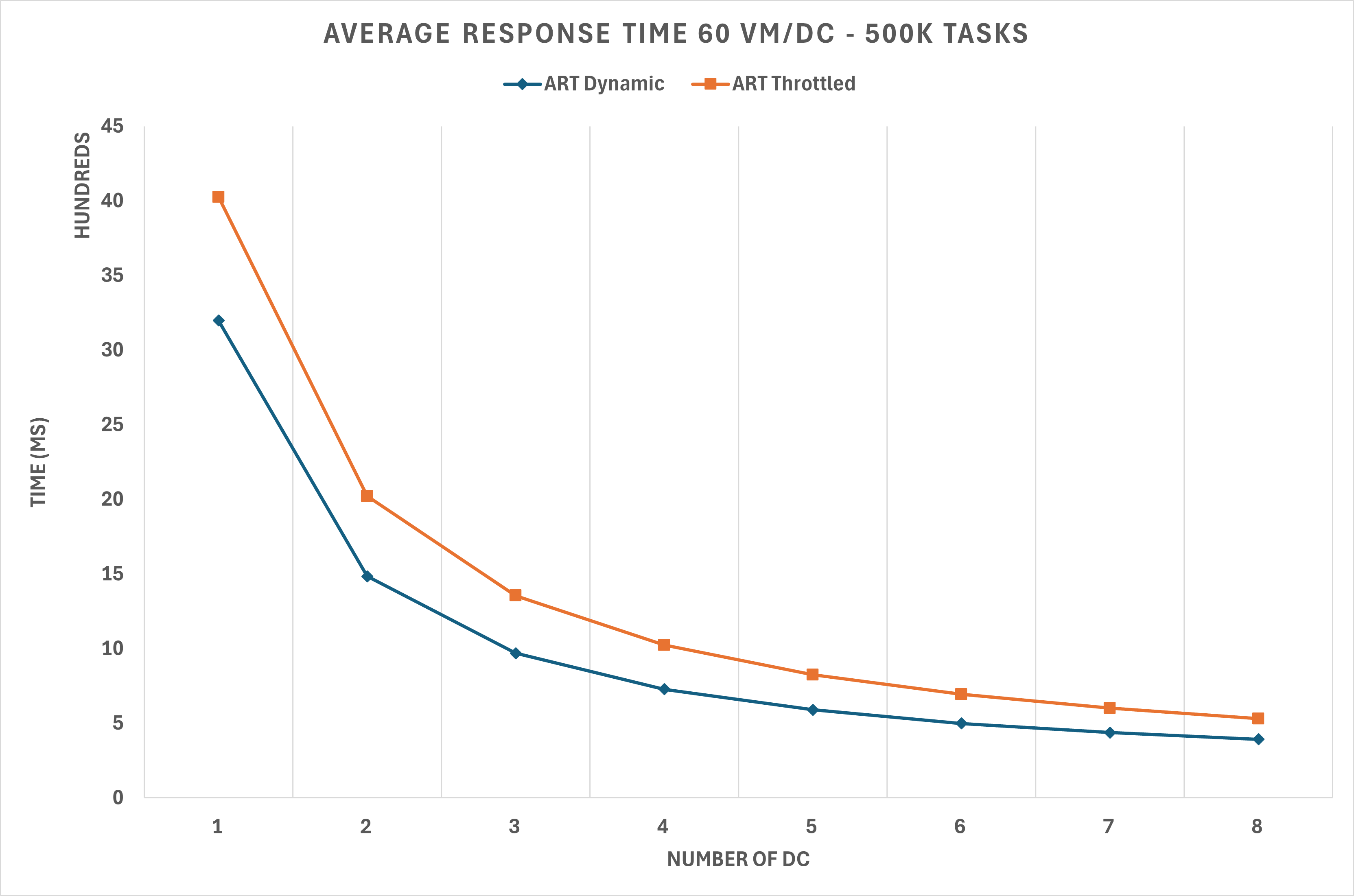}
    \caption{Average Response Time For Varying Number Of DCs Using 60VM/DC For 500k Tasks}
    \label{fig:art60vm}
\end{figure}

\begin{figure}[t]
    \centering
    \includegraphics[width=\columnwidth]{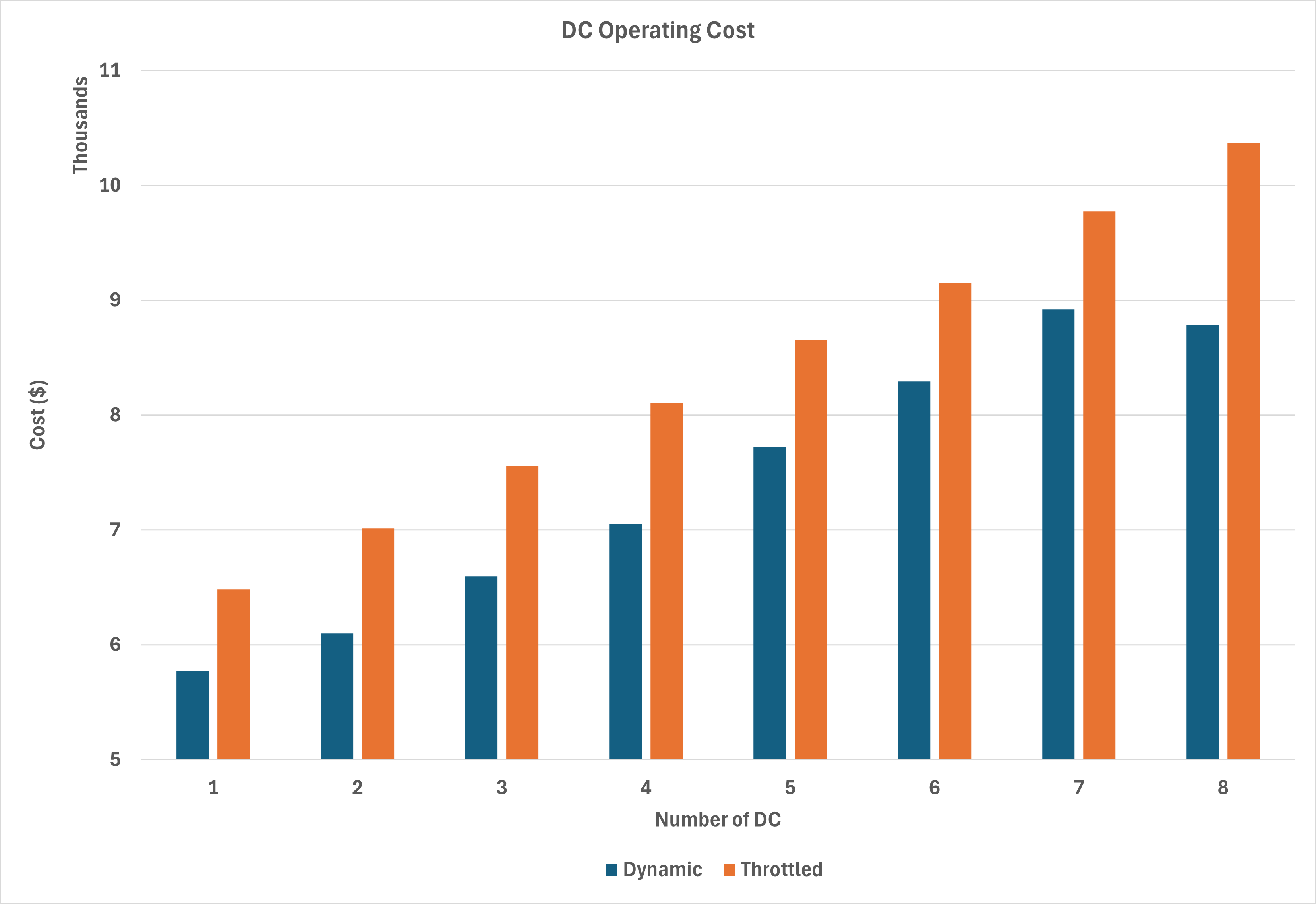}
    \caption{Data Center Operating Costs for 500K Tasks}
    \label{fig:dccost}
\end{figure}

\subsection{S-2: Varying DCs}
This experiment assessed the scalability and efficiency of the SBDLB algorithm by varying the number of data centers (1 to 8) while keeping 60 VMs per center, as identified optimal in Scenario 1. Key metrics included average response time, processing time, and operating costs.

As shown in Figure \ref{fig:art60vm}, SBDLB consistently outperformed throttled load balancing across all configurations. For a 500K task load, it achieved a 37\% lower average response time, with a highly significant p-value of 3.35 $\times$ 10$^{-12}$. Notably, SBDLB maintained better performance even with fewer data centers. At 500K tasks, it completed processing in ~970 ms using 3 DCs, while throttled required 4 DCs and still had a slower response time of 1024 ms. This trend held across all workloads from 100K to 500K, highlighting SBDLB’s efficiency in reducing resource use and operational overhead.

\begin{figure}[t]
    \centering
    \includegraphics[width=\columnwidth]{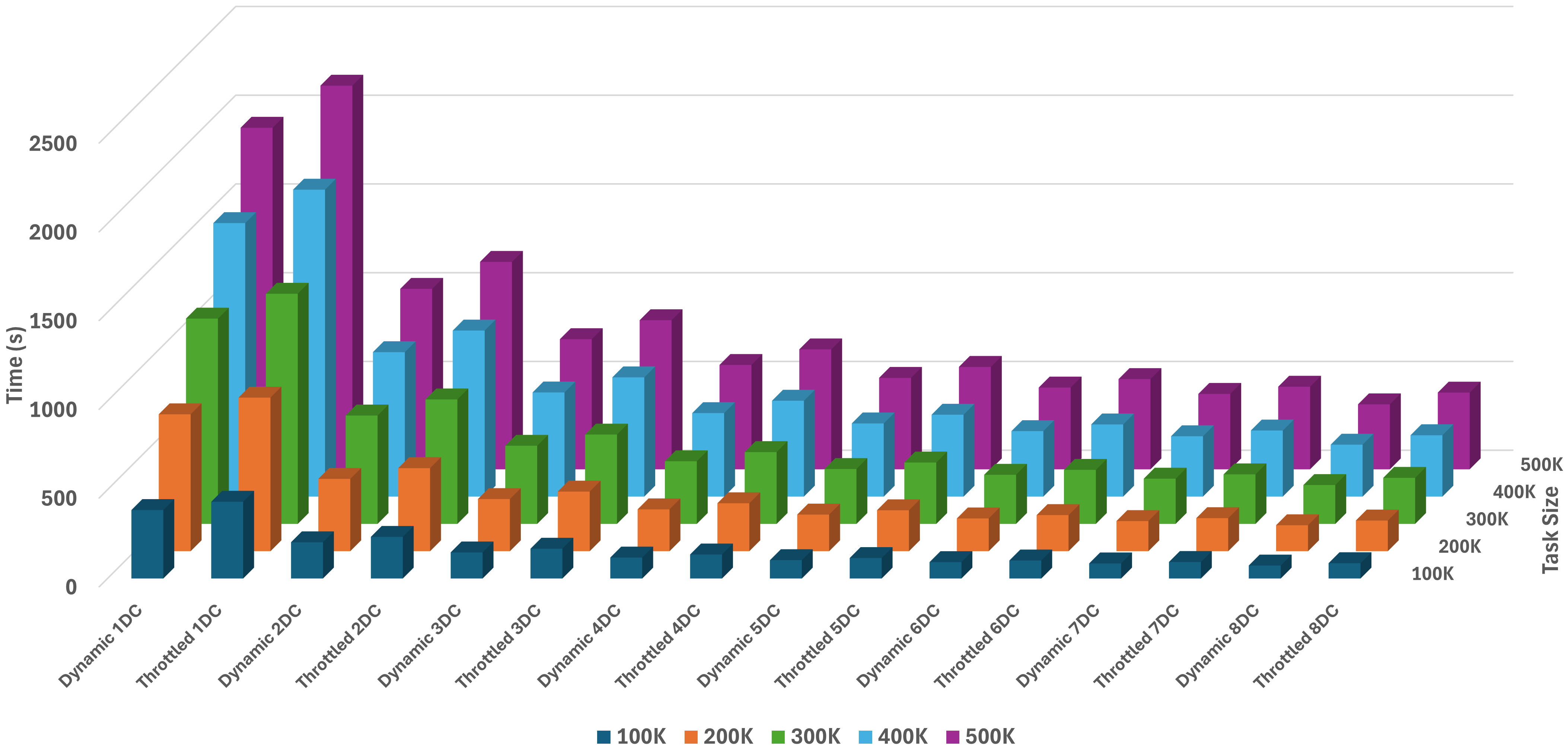}
    \caption{DC Processing Time Over 1-8 DCs And Task Size of 100K - 500K}
    \label{fig:alldcproctime}
\end{figure}

These gains translate to lower energy use, infrastructure demands, and cost—supporting sustainable, scalable cloud deployment. Figure \ref{fig:dccost} shows that SBDLB reduces operating costs, driven by faster task completion and shorter resource active time. Figure \ref{fig:alldcproctime} further shows that SBDLB cut data center processing time by 13\% on average, with a statistically significant p-value of 4.48 $\times$ 10$^{-9}$. These results reinforce SBDLB’s advantages in performance, cost-efficiency, and sustainable resource management.

\begin{figure}[t]
    \centering
    \includegraphics[width=\columnwidth]{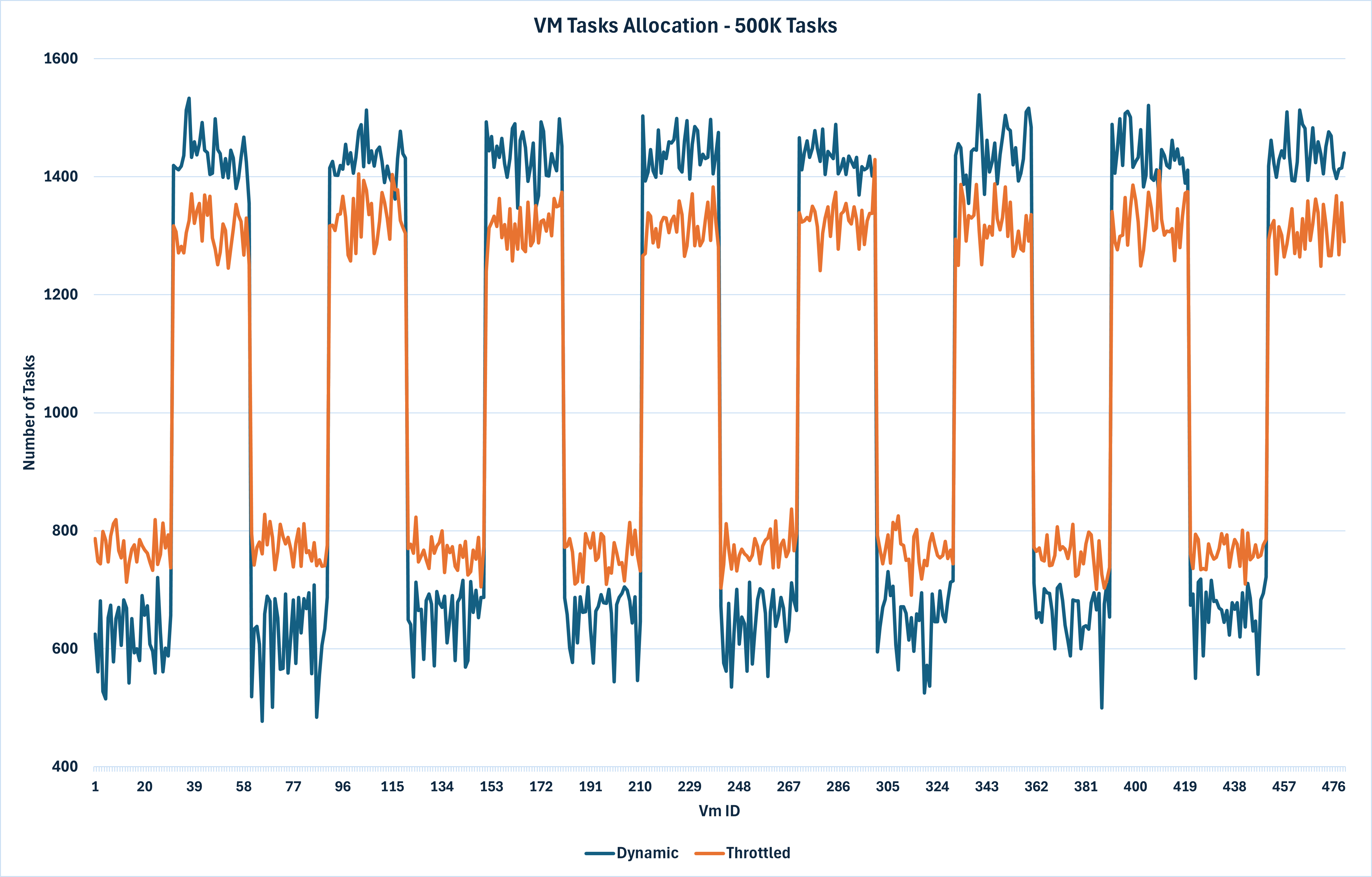}
    \caption{VM Task Distribution for 500K Tasks}
    \label{fig:vmaloc}
\end{figure}

\subsection{S-3: Task Allocation}
This experiment evaluated task distribution strategies in a heterogeneous cloud setup, showing that SBDLB achieves better performance and resource utilization than the throttled approach.

As shown in Figure \ref{fig:vmaloc}, SBDLB assigns more tasks to high-spec VMs (crests) and fewer to low-spec ones (troughs), optimizing processing power and avoiding bottlenecks. In contrast, the throttled method distributes tasks uniformly, overloading weaker VMs and underusing stronger ones. These results confirm that intelligent, capability-aware scheduling significantly improves efficiency and system performance.

\subsection{S-4: 24-Hour Variation}
To provide a more comprehensive evaluation, this study analyzes SBDLB over a 24-hour period, distinguishing between peak and non-peak hours based on user demand \cite{rajput2014simulation}, \cite{dubey2017performance}. Hourly configurations are detailed in Table \ref{tab:24hour}. To better reflect real-world workloads, the simulation introduces random variations in batch size and batch count per hour. This stochastic setup accounts for natural demand fluctuations, enhancing the robustness and realism of the results.


\begin{table}[h]
    \centering
    \caption{Batch Processing Details by Hour Type}
    \label{tab:24hour}
    \begin{tabular}{|p{1.2cm}|p{2.9cm}|p{1.7cm}|p{1.3cm}|}
        \hline
        \textbf{Hour Type} & \textbf{Hours} & \textbf{Batch Sizes} & \textbf{Total Batches} \\
        \hline
        Peak & 8-10, 13-14, 17-22 & 5K, 5.5K, 6K & 18,19,20 \\
        \hline
        Non-Peak & 0-7, 11-12, 15-16, 23-24 & 3K, 3.5K, 4K & 9,10,11 \\
        \hline
    \end{tabular}
\end{table}

\subsubsection{Average Response Time Per Hour}
\begin{figure}[t]
    \centering
    \includegraphics[width=\columnwidth]{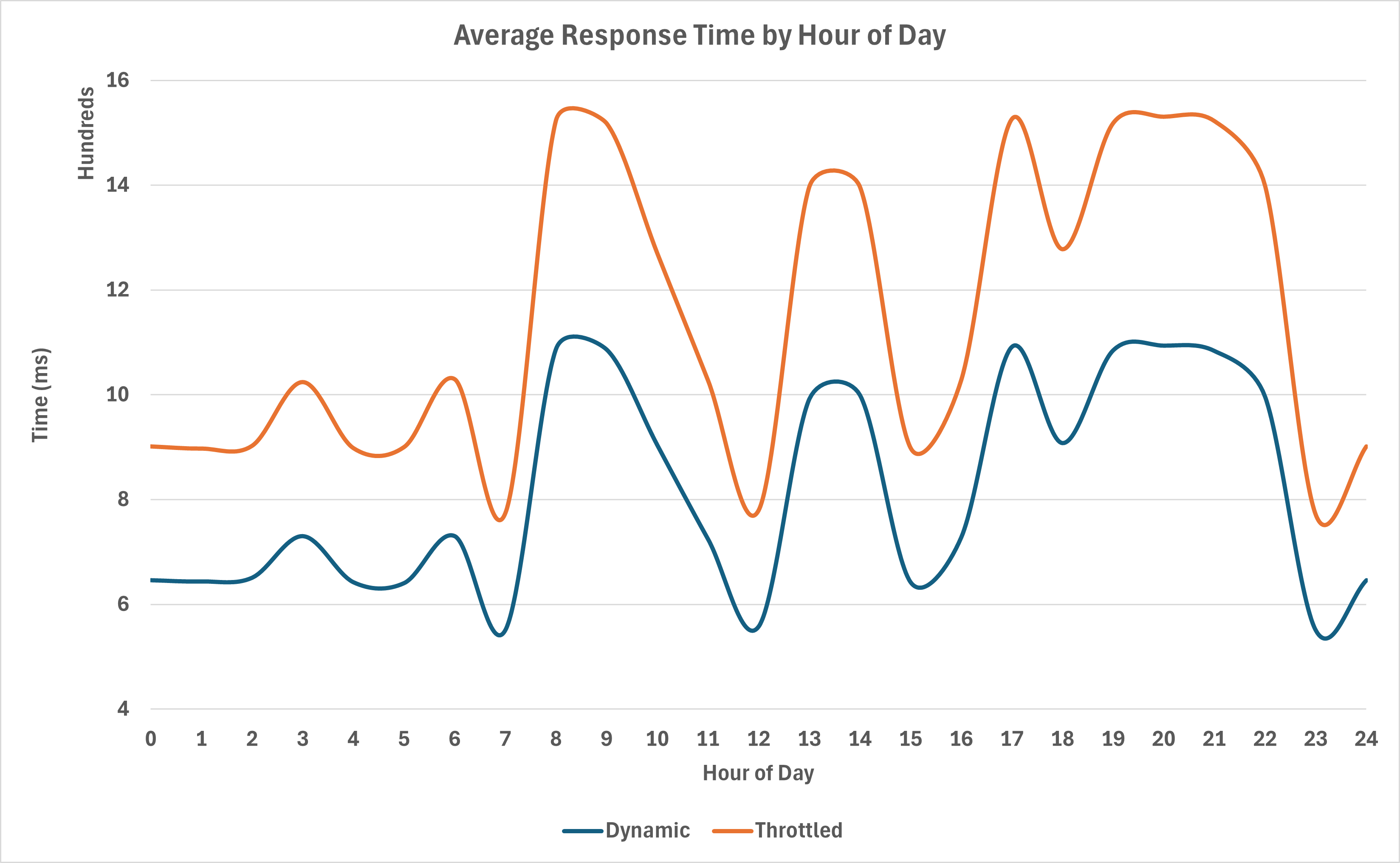}
    \caption{Average Response Time by Hour}
    \label{fig:arthourly}
\end{figure}

Figure \ref{fig:arthourly} shows hourly average response times for SBDLB and throttled load balancing under varying workloads. SBDLB consistently outperforms throttled across the 24-hour period, adapting more effectively to workload fluctuations and optimizing resource use. Throttled shows noticeable response time spikes during peak hours (8–10 AM, 1–2 PM, and 5–10 PM), reflecting its struggle with high traffic due to static resource allocation. While both methods perform better during non-peak hours, throttled still lags slightly. Though the gap narrows under lower load, SBDLB maintains a performance edge. Faster task completion with SBDLB also reduces congestion, operational costs, and energy use, emphasizing its advantages in both efficiency and sustainability.

\begin{figure}[t]
    \centering
    \includegraphics[width=\columnwidth]{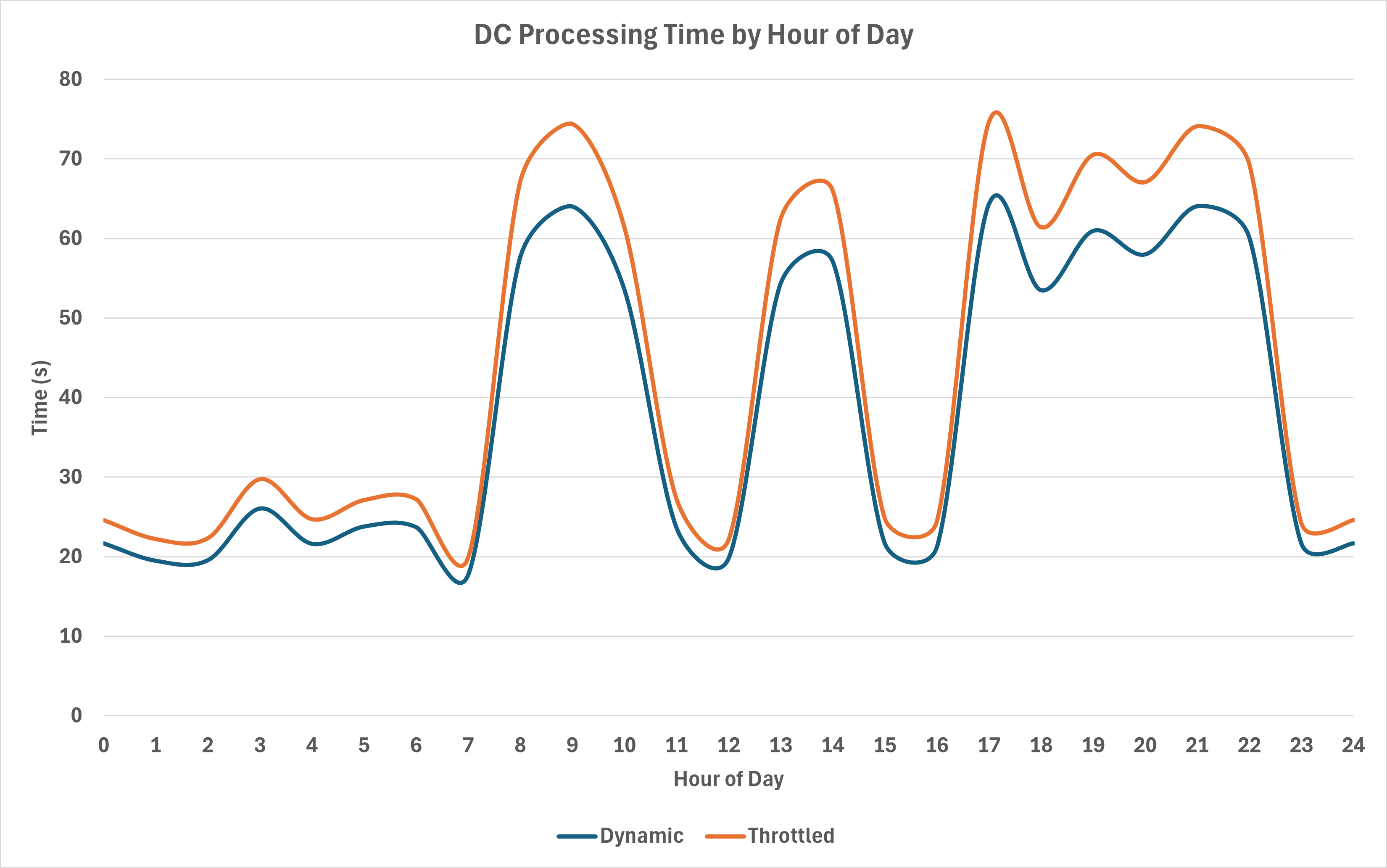}
    \caption{DC Processing Time by Hour}
    \label{fig:prochourly}
\end{figure}


\subsubsection{DC Processing Time Per Hour}
Figure \ref{fig:prochourly} shows hourly data center processing times for SBDLB and throttled load balancing. SBDLB consistently reduces processing time across all hours, handling both peak and non-peak workloads more efficiently. During peak hours, throttled struggles with workload surges, while SBDLB maintains better performance. In non-peak hours, both improve, but throttled shows greater variability, whereas SBDLB remains stable. By completing tasks faster, SBDLB not only increases efficiency, but also reduces energy use and operational costs, making it a more cost-effective cloud solution.

\subsubsection{DC Operating Cost Per Hour}
Hourly data center costs follow the trend in Figure \ref{fig:prochourly}, as cost is proportional to processing time. Both metrics peak during high-load hours and decline during non-peak periods. SBDLB consistently reduces costs compared to throttled, with the greatest savings during peak hours. In non-peak times, the cost difference narrows but still favors SBDLB.

\subsection{Total Cost Analysis Over 24 Hours}
Operating the data center for 24 hours (per Table \ref{tab:24hour}) costs \$22,818 with SBDLB versus \$26,246 with throttled—a 15.02\% reduction. This cut is significant for both cost and environmental impact. By optimizing processing during peak hours, SBDLB lowers energy use, helping reduce carbon emissions from data centers—a major source of global energy demand \cite{patel2024modeling}. Efficient load balancing not only saves money but also supports sustainability by reducing computational overhead and the data center’s carbon footprint.

\section{Conclusion}
Cloud technology has become central to modern digital infrastructure, with global adoption on the rise. Efficient load balancing is vital for optimizing performance, enhancing user experience, and reducing operational costs. This study proposes SBDLB, a dynamic load balancing method that adapts to varying conditions and outperforms the traditional throttled approach. Results show SBDLB reduces task response time by 34\%–37\%, lowers data center processing time by 13\%, and completes workloads more efficiently using fewer resources. These gains translate to cost savings and lower energy consumption, promoting both economic and environmental sustainability. Future work will explore scalability across diverse data center setups and VM types, with potential enhancements through heterogeneous VMs and reinforcement learning for self-optimizing performance.



\bibliographystyle{IEEEtran}
\bibliography{references}

\end{document}